\documentclass[twocolumn,superscriptaddress, a4paper, accepted=2022-09-16]{quantumarticle}
\pdfoutput=1
\usepackage{amsmath}
\usepackage{amssymb}
\usepackage{graphicx}
\usepackage{hyperref}
\usepackage{physics}
\usepackage{booktabs}
\usepackage{xcolor}
\usepackage{xspace,slashed}
\usepackage{qcircuit}
\usepackage{multirow}
\usepackage[utf8]{inputenc}
\usepackage{subfigure}
\usepackage[frozencache]{minted}
\setminted[python]{
    breaklines=true,
    encoding=utf8,
    fontsize=\footnotesize,
    frame=lines
}

\usepackage{array}
\usepackage[numbers,sort&compress]{natbib}

\begin{document}

\title{Quantum simulation with just-in-time compilation}


\newcommand{\MIaff}{TIF Lab, Dipartimento di Fisica, Universit\`a degli Studi di
  Milano and INFN Sezione di Milano, Milan, Italy.}

\newcommand{\TII}{Quantum Research Centre, Technology Innovation Institute, Abu Dhabi, UAE.}

\newcommand{\CERNaff}{CERN, Theoretical Physics Department, CH-1211
  Geneva 23, Switzerland.}

\author{Stavros Efthymiou}
\affiliation{\TII}
\author{Marco Lazzarin}
\affiliation{\TII}
\author{Andrea Pasquale}
\affiliation{\TII}
\affiliation{\MIaff}
\author{Stefano Carrazza}
\affiliation{\MIaff}
\affiliation{\CERNaff}
\affiliation{\TII}

\begin{abstract}
    Quantum technologies are moving towards the development of novel hardware
    devices based on quantum bits (qubits). In parallel to the development of
    quantum devices, efficient simulation tools are needed in order to design
    and benchmark quantum algorithms and applications before deployment on
    quantum hardware. In this context, we present a first attempt to perform
    circuit-based quantum simulation using the just-in-time (JIT) compilation
    technique on multiple hardware architectures and configurations based on
    single-node central processing units (CPUs) and graphics processing units
    (GPUs). One of the major challenges in scientific code development is to
    balance the level of complexity between algorithms and programming
    techniques without losing performance or degrading code readability. In this
    context, we have developed {\tt qibojit}: a new module for the Qibo quantum
    computing framework, which uses a just-in-time compilation approach through
    Python. We perform systematic performance benchmarks between our JIT
    approach and a subset of relevant publicly available libraries for quantum
    computing. We show that our novel approach simplifies the complex aspects of
    the implementation without deteriorating performance.
\end{abstract}

\maketitle

\section{Introduction}

The growing interest in quantum technologies for computational tasks which could
exceed classical devices performance has received a boost thanks to the
availability of noisy intermediate-scale quantum (NISQ) devices~\cite{nisq} and
recent promising results~\cite{supremacy, zhong2020quantum}. We observe
important steps towards the development of stable and efficient quantum
processing units (QPUs), following the gate-based model of quantum
computation~\cite{google,ibmq,rigetti,intel} or quantum
annealing~\cite{dwave,dwaveneal}.

Despite the effort in QPU technology development, aspects involving theory and
modeling do still require classical simulation of quantum computing to develop
new algorithms and applications. High performance quantum simulation serves as a
testing and profiling tool for the development of quantum algorithms, while from
an experimental point of view it provides a reference for benchmarks and error
simulation.

\begin{figure}
    \centering
    \includegraphics[width=\columnwidth]{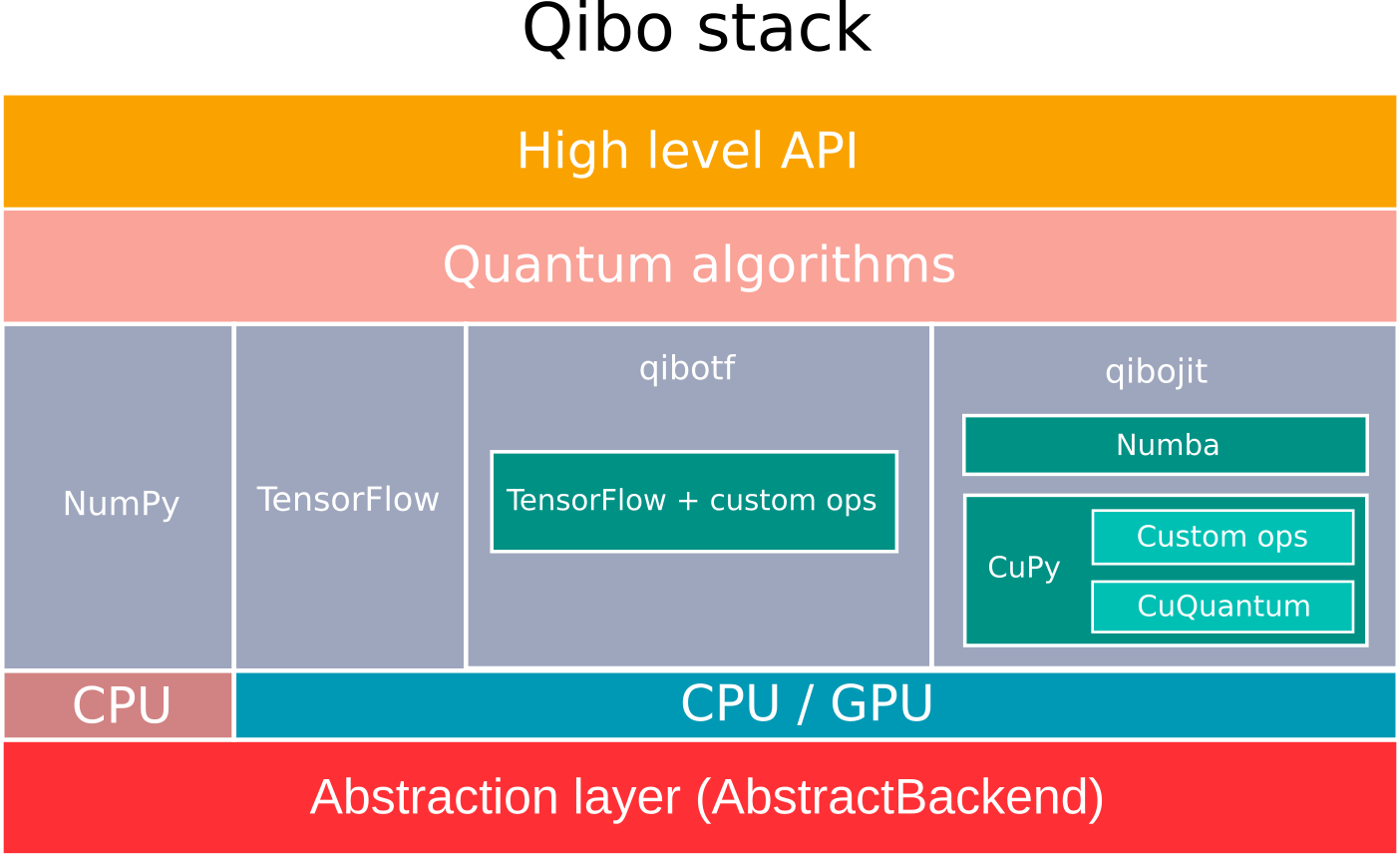}
    \caption{Schematic view of the Qibo structure design.}
    \label{fig:stack}
\end{figure}

In quantum computing the state $\psi$ of a system of $n$ qubits is represented
by a vector of $2^n$ complex probability amplitudes in the computational basis.
In Schr\"odinger's approach of quantum
simulation~\cite{feynmanhybrid,chen2018classical}, each gate is applied to the
state via the following matrix multiplication
\begin{equation}\label{eq:gateapplication}
    \psi'(\sigma_1, \ldots, \sigma_n) = \sum _{\boldsymbol{\tau'}} G(\boldsymbol{\tau}, \boldsymbol{\tau'})\psi(\sigma_1,\ldots,\boldsymbol{\tau'},\ldots,\sigma_n)
\end{equation}
where the gate targeting $n_{\rm tar}$ qubits is represented by the $2^{n_{\rm
tar}}\times2^{n_{\rm tar}}$ complex matrix
$G(\boldsymbol{\tau},\boldsymbol{\tau'})=G(\tau_1,\ldots,\tau_{n_{\rm
tar}},\tau_1',\ldots,\tau_{n_{\rm tar}}')$ and $\sigma _i, \tau _i\in
\{0,1\}$.
The numerical solution to Eq.~\ref{eq:gateapplication} requires the manipulation
of state vectors of size $2^n$, which scales exponentially with the number of
qubits, and the subsequent linear algebra operations related to the application
of unitary gates. Thus, quantum simulation tools on classical hardware need to
take into account both challenges and provide efficient solutions.

In this context, we have developed the
Qibo~\cite{qibo2021,stavros_efthymiou_2021_5711842,Carrazza:2022dck} framework,
an open-source, full stack API written in Python, which supports circuit-based
quantum simulation, adiabatic evolution simulation and quantum hardware
control~\cite{qibolab2022}.
The Qibo structure since release 0.1.7 is shown in Fig.~\ref{fig:stack}. The
high-level API and pre-coded quantum algorithms are implemented following a
backend agnostic approach. Each backend provides specialized methods to achieve
maximum performance on multiple devices, including hardware accelerators, such
as multi-threading CPU, GPU and multi-GPU configurations.

The main disadvantage associated with the development and maintenance of
scientific software with parallel computing and hardware acceleration support is
the need to maintain a large code-base of algorithms defined in compiled
languages (such as Fortran, C++ and CUDA). This requires a non-negligible
level of programming experience for the developer.
Furthermore, testing and deployment of these codes requires custom workflows
which should build pre-compiled binaries for a target subset of platforms and
architectures.

To address these issues, we published the {\tt qibojit}
backend~\cite{stavros_efthymiou_2021_5248470} which supports efficient
circuit-based quantum simulation through just-in-time
(JIT)~\cite{lam2015numba,cupy_learningsys2017} compilation, with Python as
input programming language interface.
The Python JIT approach provides to the code developer the possibility to
maintain a modern project layout, with automatic documentation, testing
workflows and standard deployment procedure with minor changes to the
algorithmic part of the code. The code readability and homogeneity are
preserved, and it makes the installation on different platforms easier. In this
paper we first present the layout adopted by {\tt qibojit} and then perform a
systematic benchmark to quantify the impact on performance for quantum computing
tasks.

Finally, it is important to highlight that similar quantum simulators are
implemented by other research collaborations and companies. Some examples
included in the benchmark section of this work are Qiskit~\cite{qiskit} from
IBM, Cirq and
qsim~\cite{cirq_developers_2021_5182845,quantum_ai_team_and_collaborators_2020_4023103}
from Google, ProjectQ~\cite{projectq_1,projectq_2} by ETH Z\"urich,
HybridQ~\cite{9651384} by NASA, Qulacs~\cite{qulacs} and
QCGPU~\cite{kelly2018simulating}.

The paper is organized as follows. In Sec.~\ref{sec:methodology} we present the
technical details of the {\tt qibojit} implementation as a module for the Qibo
framework, highlighting the code design and structure. The
Sec.~\ref{sec:benchmarks} presents performance benchmarks of all Qibo backends as
well as other quantum simulators. Finally, in Sec.~\ref{sec:conclusion} we
present our conclusion and outlook.

\section{Methodology}
\label{sec:methodology}

\begin{figure}
    \centering
    \includegraphics[width=0.45\textwidth]{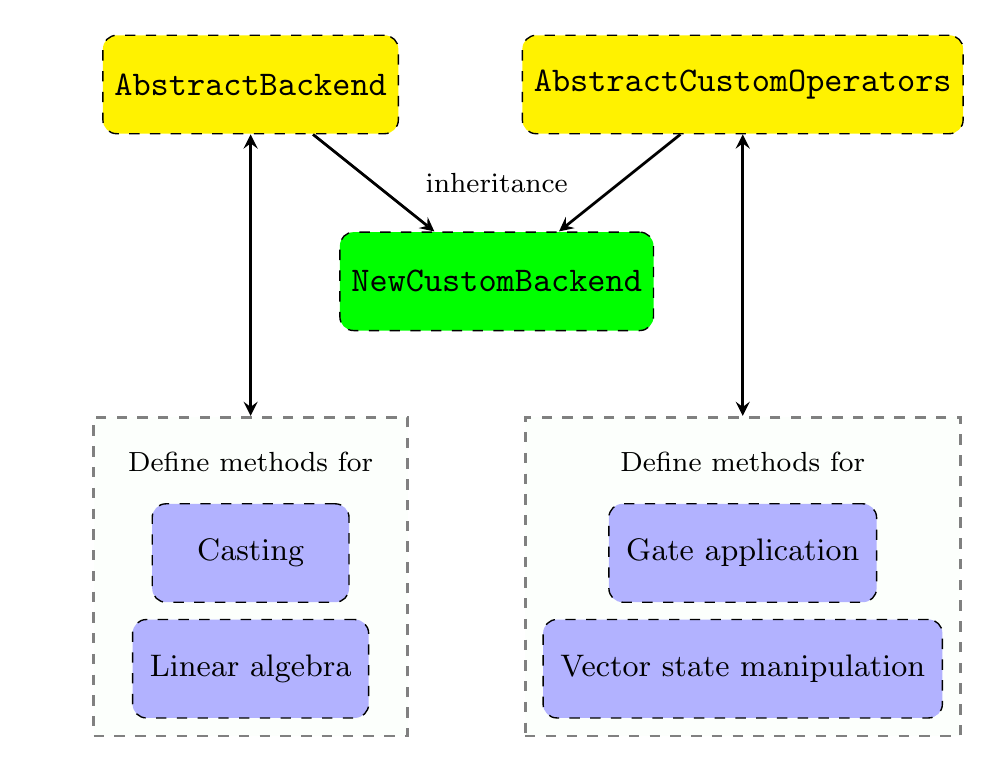}
    \caption{Flowchart describing how to implement a new custom backend.}
    \label{fig:flowchart}
\end{figure}

Qibo provides multiple backends for implementing the matrix multiplication of
Eq.~\ref{eq:gateapplication} which are based on different technologies including
pre-compiled binaries and just-in-time compilation as shown in
Fig.~\ref{fig:stack}. All backends inherit from the {\tt AbstractBackend} class
and define its properties and methods using primitives provided by Python
libraries, such as NumPy~\cite{numpybook}, and custom operations coded in Python
or low-level languages such as C++ and CUDA.
The abstract methods include general algebraic and linear algebra operations,
such as element wise vector operations, tensor products, eigenvalue and
eigenvector methods, as well as specialized operations for applying gates to
state vectors and density matrices, following Eq.~\ref{eq:gateapplication}. This
abstraction layer allows us to disentangle the core Qibo code and applications
from a specific backend or Python library. Furthermore, it provides a layout for
users that would like to define a new backend that is compatible with Qibo. This
can be done just by inheriting from the {\tt AbstractBackend} and the {\tt
AbstractCustomOperator} and defining their abstract methods as shown in
Fig.~\ref{fig:flowchart}.

The basic backends included in the Qibo package, distributed from PyPI~\cite{pypi_qibo}
and conda-forge~\cite{conda_qibo}, are {\tt numpy} and {\tt tensorflow}.
These backends implement basic algebraic operations using primitives of the
underlying library, NumPy~\cite{numpybook} and
TensorFlow~\cite{tensorflow2015-whitepaper}. The specialized operations for
applying gates are based on the {\tt einsum} method which is exposed as a
primitive in both libraries. This provides satisfactory performance when
simulating circuits up to 20 qubits. The {\tt numpy} backend is available by
default when installing Qibo and it is designed to support a high number of
architectures, including {\tt arm64}, and thus be deployed in multiple contexts,
including laboratory devices. The optional {\tt tensorflow} backend provides moderate
performance and the possibility to perform automatic differentiation which is
useful for quantum machine learning applications.

To efficiently simulate circuits with a larger number of qubits, we extend
these basic backends with custom operators. In particular, {\tt tensorflow} is
extended by {\tt qibotf}~\cite{qibotf} and {\tt numpy} is extended
by {\tt qibojit}~\cite{jiturl}.
These are not included in the basic Qibo library but can be installed
as separate Python packages.
These backends keep using their parent libraries (NumPy and TensorFlow) for
basic algebraic and linear algebra operations, however the gate application
methods are replaced by custom operators. Unlike the {\tt einsum} approach,
which duplicates the state vector while applying a gate,
custom operators perform in-place updates.
This reduces both memory requirements and execution time since custom operators
modify directly the initial state vector based on the gates applied.
Furthermore, the custom operators exploit the sparsity of matrices
associated with some common operations, such as Pauli gates and controlled
gates, to reduce the number of operations required to apply each gate. In particular,
if the application of a specific gate modifies only a few components of the initial state,
using custom operators we update directly these particular elements,
avoiding the matrix multiplication of Eq.~\ref{eq:gateapplication}.

The basic custom operator defines the application of an arbitrary single-qubit
gate to a state vector. An example of this operator for the {\tt qibojit}
backend is shown below.
\begin{minted}[fontsize=\scriptsize]{python}
from numba import njit, prange

@njit(parallel=True, cache=True)
def apply_gate_kernel(state, gate, target):
    """Operator that applies an arbitrary one-qubit gate.

    Args:
        state (np.ndarray): State vector of size (2 ** nqubits,).
        gate (np.ndarray): Gate matrix of size (2, 2).
        target (int): Index of the target qubit.
    """
    k = 1 << target
    # for one target qubit: loop over half states
    nstates = len(states) // 2
    for g in prange(nstates):
        # generate index with fast binary operations
        i1 = ((g >> m) << (m + 1)) + (g & (k - 1))
        i2 = i1 + k
        state[i1], state[i2] = (gate[0, 0] * state[i1] + \
                                gate[0, 1] * state[i2],
                                gate[1, 0] * state[i1] + \
                                gate[1, 1] * state[i2])
    return state
\end{minted}
Additional operators that follow a similar approach are used to apply gates with
more target qubits, as well as controlled gates. All these operators take
advantage of multi-threading CPUs and GPUs by parallelizing the loop over state
elements, the cost of which scales exponentially with the number of qubits.
Furthermore, we provide specialized operators for applying Pauli X, Y and Z
gates and the SWAP gate, which use more simplified kernels inside the loop. In
order to simulate real measurements, we provide a custom operator for collapsing
and re-normalizing states and a method for sampling shot frequencies based on
Metropolis algorithm~\cite{Metropolis:1953am}. Both {\tt qibotf} and {\tt
qibojit} define the same custom operators but use different technologies to
interface them with Python. These are analyzed in what follows.

In {\tt qibotf} we use TensorFlow custom operators written in C++ and CUDA.
These need to be compiled before the execution, a step that typically improves performance
but could complicate installation and make it very device specific.
Nevertheless, this is the first custom backend released for Qibo and includes multi-threading
CPU, GPU and multi-GPU support.

The latest backend added to Qibo is {\tt qibojit},
which implements custom operators based on a just-in-time compilation approach.
{\tt qibojit} also supports multi-threading CPU, GPU and multi-GPU configurations.

For CPU we write operators in Python using Numba's~\cite{lam2015numba} {\tt njit} decorator with
a set of signatures for each function that specify both return and argument types.
This decorator compiles the Python code using LLVM.
Moreover, the loop over the state elements is parallelized using Numba's
\texttt{numba.prange} method. To further speed up the circuit execution the appropriate
indices for each update are generated on-the-fly using fast binary operations.

For GPU, we choose Cupy~\cite{cupy_learningsys2017} as the main driver, which
enables us to follow the CPU approach based on on-the-fly compilation. We also
tried different GPU backends, including Numba and Jax~\cite{jax2018github} for
Python or in C++ Eigen3~\cite{eigenweb}, ViennaCL~\cite{doi:10.1137/15M1026419}
and NVIDIA thrust~\cite{thrust}. The main problems with these options concern
the lack of linear algebra operations and the difficulties in writing custom
operators. We decided not to choose Numba because we observe a significant
overhead when simulating circuits with a small number of qubits.
The implementation of the custom operators in the Cupy backend was performed using the
{\tt RawKernel} method, which allows us to define custom CUDA kernels written in C++,
which are compiled using nvcc~\cite{nvcc} at their first invocation and cached for each device.
This method also takes care of exposing these compiled kernels to Python.
Another positive aspect of Cupy is the compatibility with AMD ROCm,
which enables us to run Qibo on setups with ROCm-compatible GPUs.

We also provide a different GPU simulator, within the {\tt qibojit} backend,
which is based on cuQuantum~\cite{cuquantum}, a quantum simulation library from
NVIDIA. Its addition to Qibo was facilitated since the main driver is Cupy which
is already employed by {\tt qibojit}.
This backend replaces the custom kernels with primitives from the cuQuantum
library. The main advantage is the fact that we no longer need to write C++ or
CUDA code to achieve good performances with large number of qubits. However,
using an external library instead of custom operators comes at the cost of
having less control over the code and there can also be some missing features
that need to be included manually. In our case, if a particular operator is not
defined in the cuQuantum library, the compatibility with Cupy allows us to fall
back to the custom operators of the Cupy backend to maintain good performances
without complicating the code.

For a full list of primitives and models for quantum computing simulation
available in Qibo 0.1.7 please refer to~\cite{qibo2021} and Sec.~3 in~\cite{Carrazza:2022dck}.

\section{Benchmarks}
\label{sec:benchmarks}

\begin{figure*}
    \centering
    \includegraphics[width=0.75\textwidth]{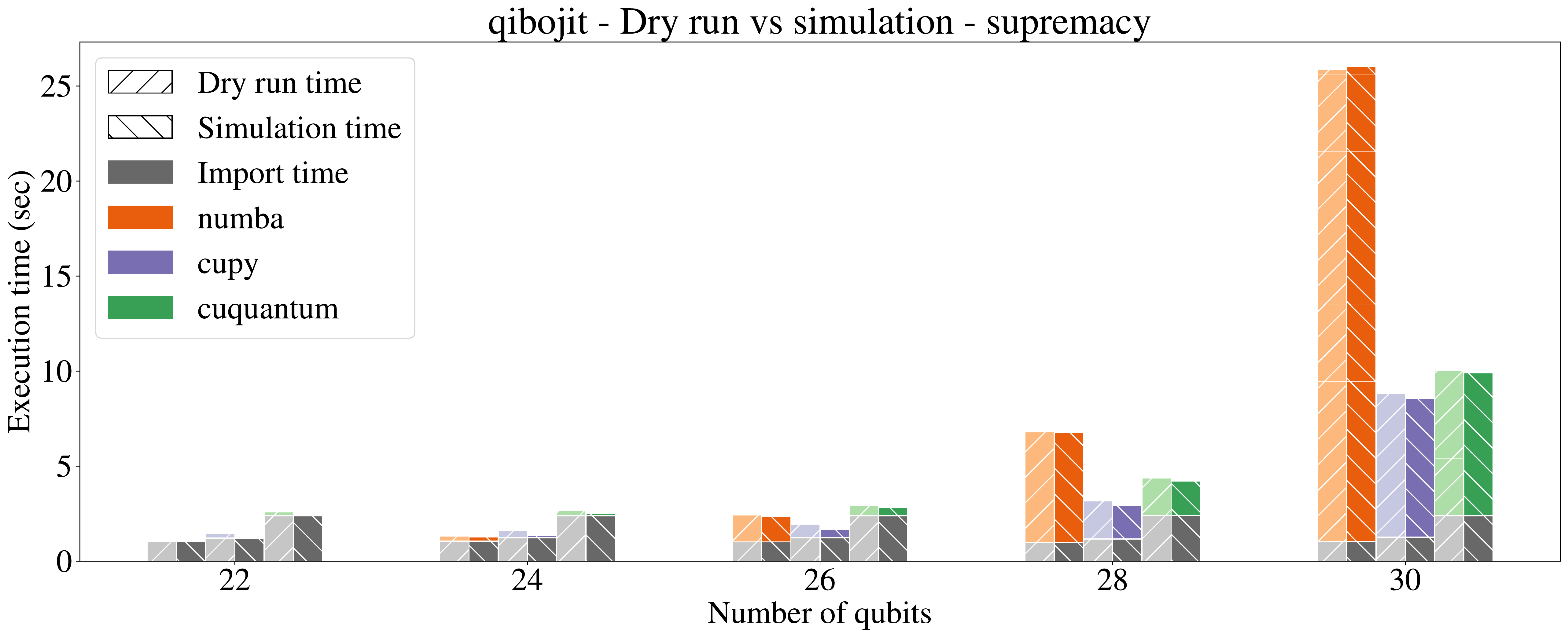}
    \caption{Comparison between import, dry run and simulation times for the three platforms of the {\tt qibojit} backend.}
    \label{fig:qibojit_dry_vs_simulation}
\end{figure*}

In this section we compare performance of the different Qibo backends and other
open-source libraries on various tasks, including
simulation of quantum circuits as well as adiabatic time evolution. The
benchmarks were performed with an AMD EPYC 7742 CPU with 128 threads and
2TB of RAM and a NVIDIA RTX A6000 GPU with 48GB of memory, unless otherwise
noted. Qibo was installed in a Python 3.9 conda environment with the
dependencies shown in Table~\ref{tab:qibo_versions}. The source code used to
generate the results in this section is publicly available in the following
repository~\cite{stavros_efthymiou_2022_6363155}.

\begin{table}
\centering
\begin{tabular}{lccc}
    \toprule
    \textbf{Name} & \textbf{Version} & \textbf{Distribution} \\
    \midrule
    qibo & 0.1.7 & pip \\
    qibojit & 0.0.4 & pip \\
    qibotf & 0.0.6 & pip \\
    tensorflow & 2.8.0 & pip \\
    numba & 0.55.1 & pip \\
    cudatoolkit & 11.6.0 & conda \\
    cupy & 10.1.0 & conda \\
    cuquantum & 0.1.0.30 & conda-forge \\
    cuquantum-python & 0.1.0.0 & conda-forge \\
    \bottomrule
\end{tabular}
\caption{Versions of Qibo and its dependencies used in the benchmarks.}
\label{tab:qibo_versions}
\end{table}

\subsection{Circuit Simulation}
\label{sec:circuit_benchmarks}

The quantum circuits used in our benchmarks are shown in
Table~\ref{tab:benchmark_circuits}. All circuits are defined using the
OpenQASM~\cite{2017arXiv170703429C,cross2021openqasm} language and ported to
each simulation library. Some libraries allow importing circuits directly from
OpenQASM, while for other libraries we coded the parsing manually. The qft,
variational and bv circuits are defined directly in OpenQASM using the
corresponding gates. The supremacy circuit is created using Cirq~\cite{github_cirq_supremacy}
and the qv circuit using Qiskit~\cite{quantum_volume_qiskit}
and are both ported to OpenQASM in order to be converted to the different
libraries.

\begin{table*}
\centering
\begin{tabular}{lcccccc}
    \toprule
    \textbf{Name} & \textbf{Notation} & \textbf{Source} & \textbf{Depth} & \textbf{Gates}
    & \textbf{Depth*} & \textbf{Gates*} \\
    \midrule
    Quantum Fourier Transform~\cite{qft} & qft & Qibo  & 60 & 480 & 58 & 450  \\
    Variational~\cite{variational} & variational & Qibo  & 4 & 90 & 2 & 30 \\
    Supremacy~\cite{supremacy, supremacy_boixo} & supremacy & Cirq  & 4 & 98 & 2 & 22\\
    Quantum Volume~\cite{quantum_volume} & qv & Qiskit  & 7 &  165 & 1 & 15\\
    Bernstein-Vazirani~\cite{bernstein_vazirani} & bv & Qibo  & 32 & 89 & 29 & 29\\
    \bottomrule
\end{tabular}
\caption{Description of circuits used in the benchmarks. The circuit depths and the number of gates
shown are referred to 30 qubits circuits. In the last two columns we show the circuit depths and the number
of gates after applying gate fusion.}
\label{tab:benchmark_circuits}
\end{table*}
\begin{figure*}[t]
    \centering
    \includegraphics[width=0.5\textwidth]{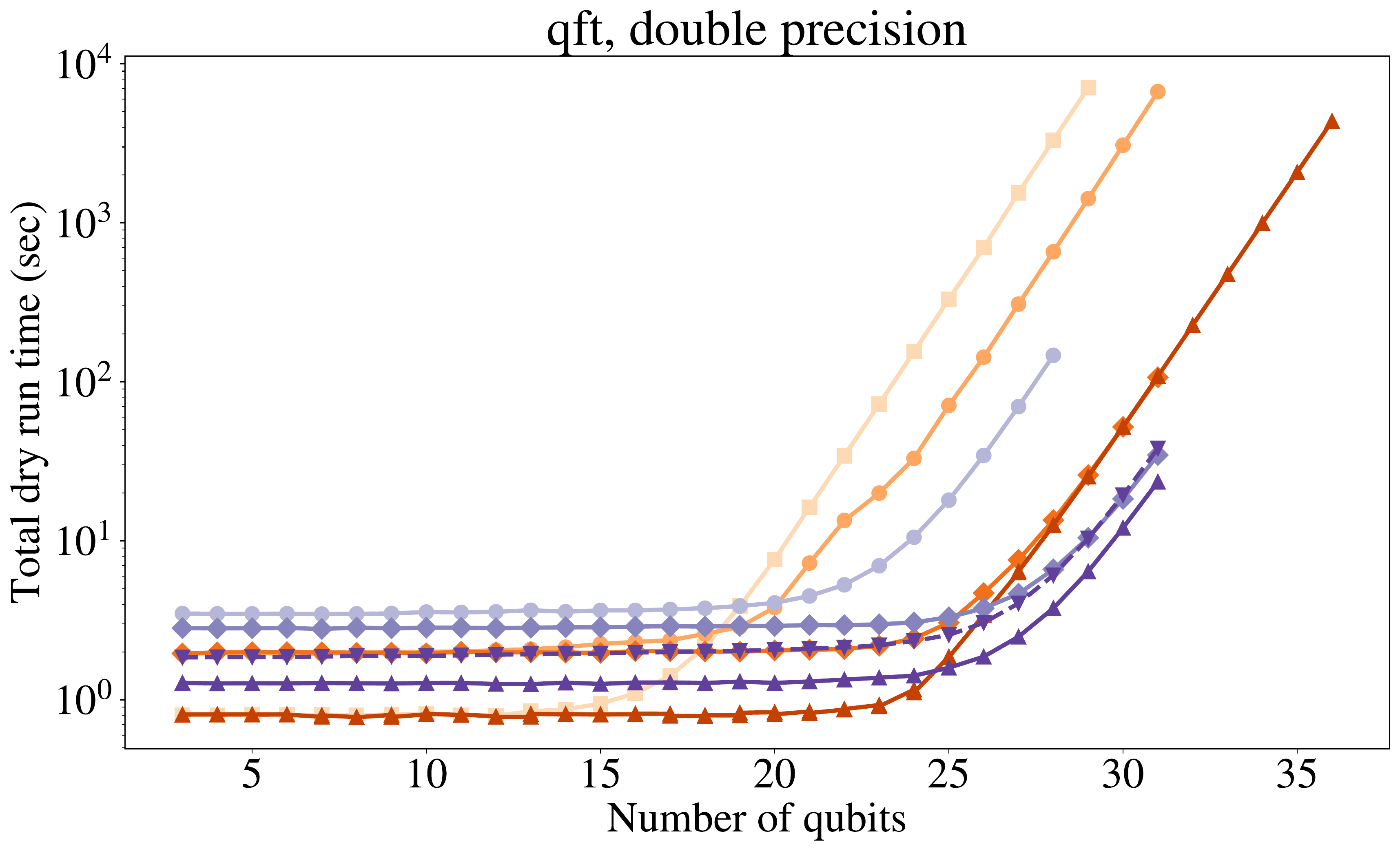}%
    \includegraphics[width=0.5\textwidth]{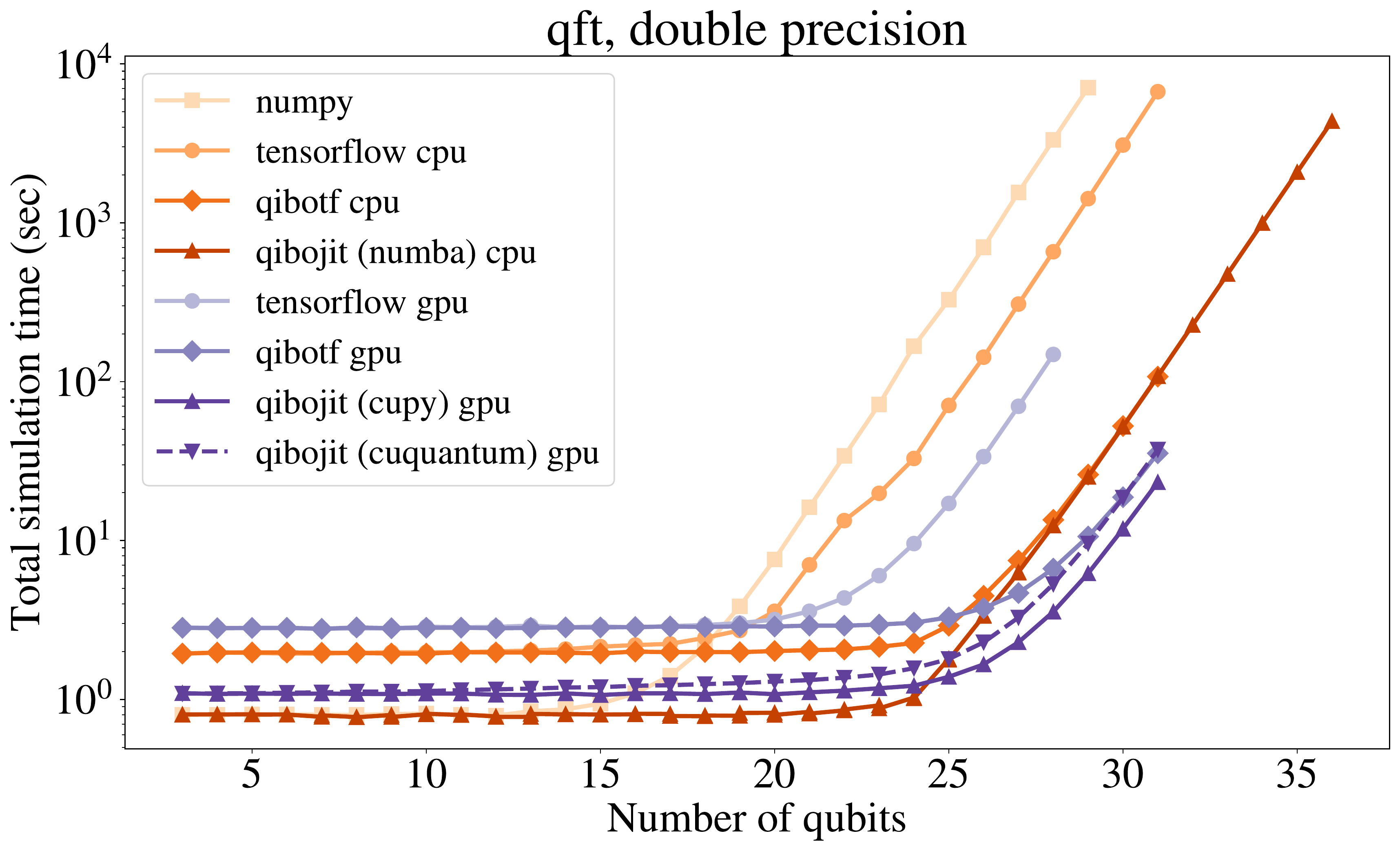}
    \caption{Total dry run (left) and simulation (right) time scaling with the number of qubits for simulating
             the qft circuit using different Qibo backends.}
    \label{fig:qft_scaling}
\end{figure*}

\begin{table*}
    \centering
    \begin{tabular}{lcccc}
        \toprule
        \textbf{backend} & \textbf{ $\Delta$m (MB)} & \textbf{m(MB)} & \textbf{dry run(s)} & \textbf{simulation(s)} \\
        \midrule
        qibojit (cupy) & 695.55 & 1093.55 & 1.92  & 0.60  \\
        qibojit (cuquantum) & 406.86 & 1804.09 & 1.035  & 0.86  \\
        qibotf (GPU) & 654.49 & 3260.28 &0.76  & 0.76  \\
        tensorflow (GPU) & 1469.14 & 4072.92 & 31.89 & 30.58 \\
        qibojit (numba) & 903.23 & 1146.02 & 2.67 & 2.57 \\
        qibotf (CPU) & 735.21 & 1276.16 & 2.52  & 2.34 \\
        tensorflow (CPU) & 4845.74 & 5385.39 & 147.01  & 146.77 \\
        numpy (CPU) & 3005.98 & 3248.69 & 697.12  & 698.65 \\
        \bottomrule
    \end{tabular}
    \caption{Memory usage, dry run and simulation times for
             different backends when simulating the qft circuit with 26 qubits.
             m denotes the maximum memory usage during the execution, while $\Delta$m represents the difference between m and the memory required for importing the dependencies.}
    \label{tab:memory}
    \end{table*}

When benchmarking libraries which involve just-in-time compilation it is
important to distinguish the first execution because it will involve a
compilation or loading of cached binaries and therefore will be slower than
subsequent executions in the same run time. In what follows, we call this first
run as \textit{dry run} and any subsequent run as \textit{simulation}.
Fig.~\ref{fig:qibojit_dry_vs_simulation} shows the difference between these two
runs for simulating the supremacy circuit using the different platforms (numba,
cupy and cuquantum) of Qibo's {\tt qibojit} backend. We observe that the
difference between the first (dry run) and second (simulation) run is negligibly
small on CPU (numba) but slightly higher on GPU (cupy, cuquantum). Note that
qibojit implements a caching algorithm for custom operators which
are generated during installation time, thus in this case negligible performance
differences between dry and simulation run-times are expected. Furthermore, a
constant of about one second is required to import the library, which can be
relevant (comparable or larger than execution time) for simulation of small
circuits. This is unlikely to impede practical usage as it is only a small
constant overhead that is independent of the total simulation load.

\begin{figure*}[t]
    \centering
    \includegraphics[width=0.5\textwidth]{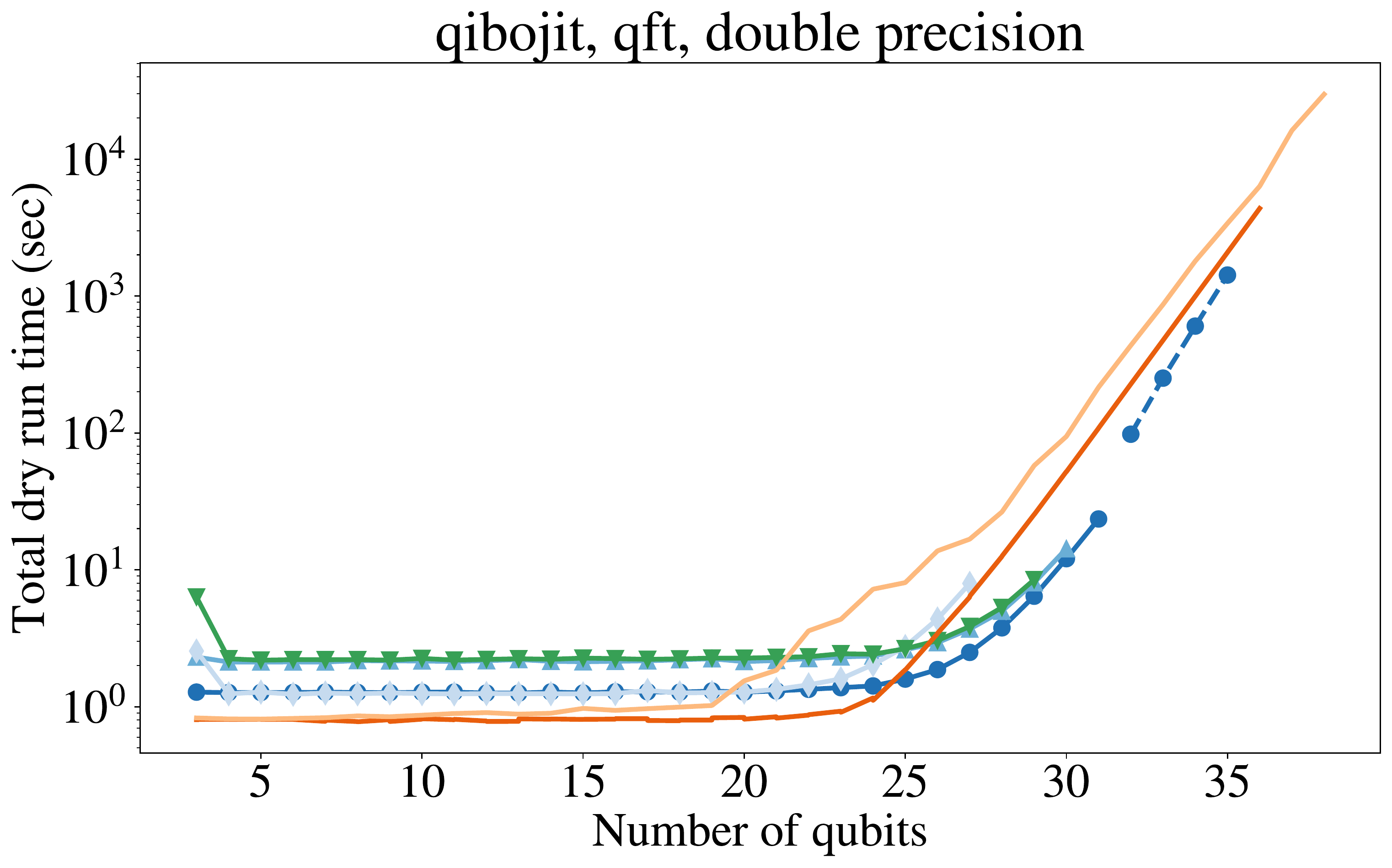}%
    \includegraphics[width=0.5\textwidth]{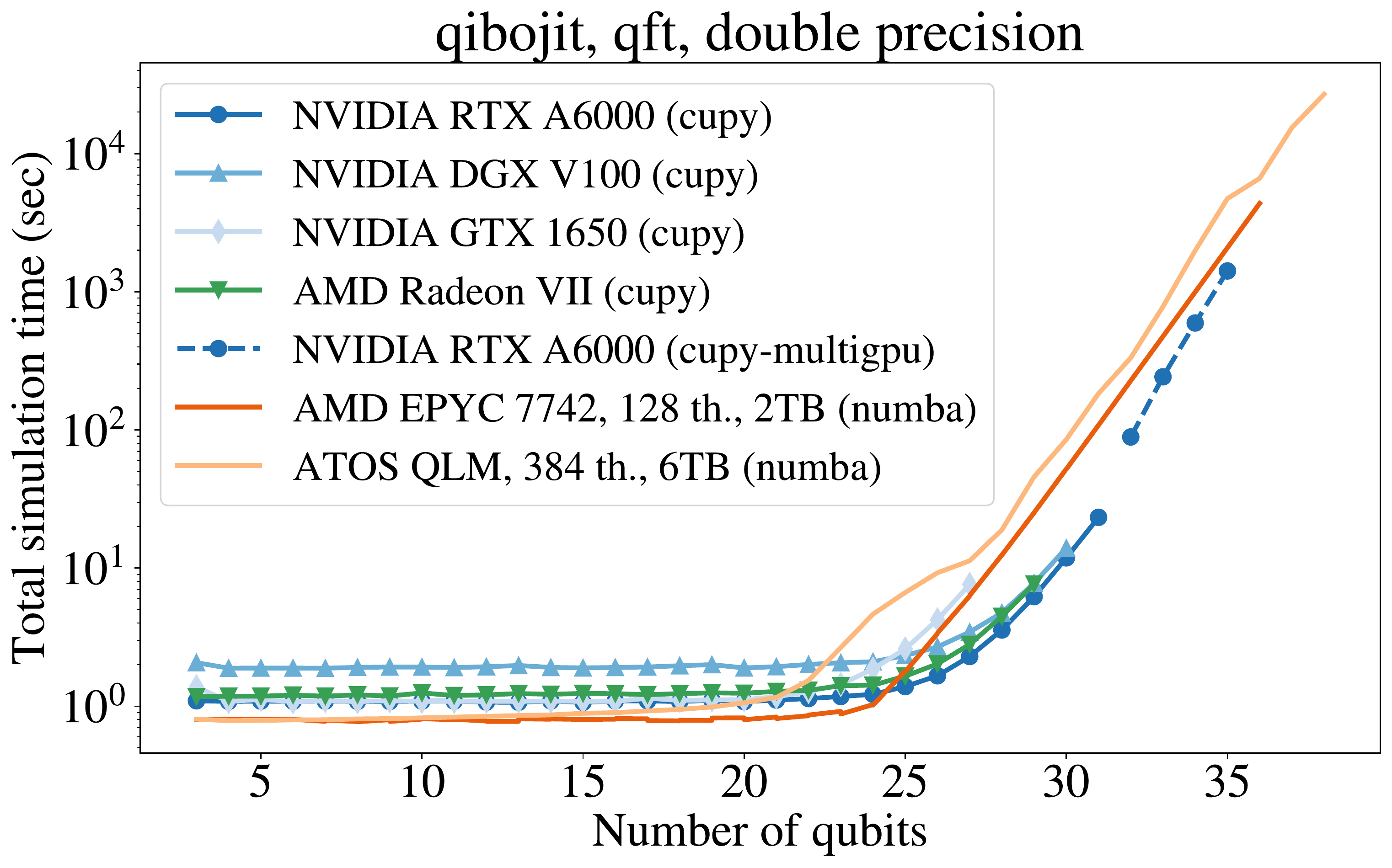}
    \caption{Total dry run (left) and simulation (right) time scaling with the number of qubits for simulating the qft circuit on different devices.}
    \label{fig:devices_scaling}
\end{figure*}

In Fig.~\ref{fig:qft_scaling} we show how the dry run (left) and simulation (right) time to execute the qft circuit
scales with the number of qubits for different Qibo backends. These plots show the total time a user would experience
when simulating the circuit, which includes the library import, allocation of the circuit and gate objects and
finally execution on the specified hardware. Up to 20 qubits this is dominated by import time and lightweight
CPU backends such as {\tt numpy} are the optimal choice. For larger circuits, the custom {\tt qibojit} and {\tt qibotf} backends
which take advantage of multi-threading CPU and GPU architectures provide a much more favorable scaling.
Moreover, {\tt qibojit} provides better performance than {\tt qibotf} despite its code simplicity.
We also note that {\tt qibojit} and {\tt qibotf} perform in-place updates, in contrast to numpy and tensorflow which duplicate
the state vector, a feature that reduces both time and memory requirements significantly.

In order to quantify the advantages associated with the JIT approach
in Table~\ref{tab:memory} we show both memory footprints and execution
times separated in dry run and simulation when executing a 26 qubits qft
circuit. We observe that the considerable reduction of the
execution times does not imply an increase in memory usage.
In fact, the backends that implement custom operators and in-place updates,
{\tt qibojit} and {\tt qibotf}, require less memory in order to perform the simulation
compared to the tensorflow and the numpy backend due to the multiple copies of the state
vector employed by these two.

In Fig.~\ref{fig:devices_scaling} we show the total dry run and simulation times
scaling with the number of qubits in the circuit, but now focusing on the {\tt
qibojit} backend and using it on different hardware configurations. As mentioned
earlier, CPU is preferable for smaller circuits due to faster import times,
which are dominating execution time at this region. We observe that {\tt
qibojit} can reach high qubit values thanks to its state vector in-place memory
updates and it can operate on multiple systems, including commercial solutions
such as ATOS QLM~\cite{atos} hardware. For circuits with more than 25 qubits the
exponential scaling starts to appear, and high-end GPUs provide an advantage.
Lower end GPUs, such as the NVIDIA GTX 1650 do not seem to provide any advantage
over a powerful CPU and their limited memory (4 GB) prohibits the simulation of
circuits with more than 27 qubits. The newest NVIDIA RTX A6000 is the fastest of
our devices. Moreover, in order to test {\tt qibojit} performance on AMD ROCm
GPUs we include numbers for the AMD Radeon VII with 16GB which confirms
competitive results.
\begin{figure*}
    \centering
    \includegraphics[width=0.41\textwidth]{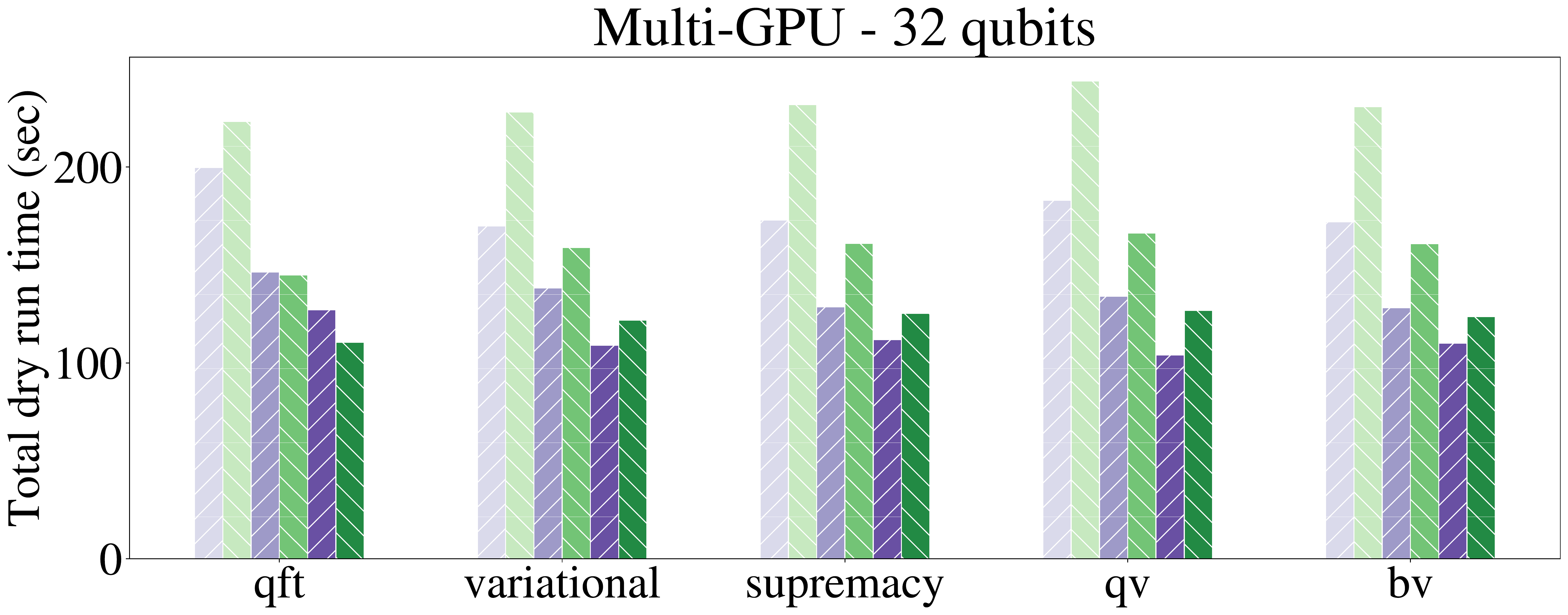}
    \includegraphics[width=0.5\textwidth]{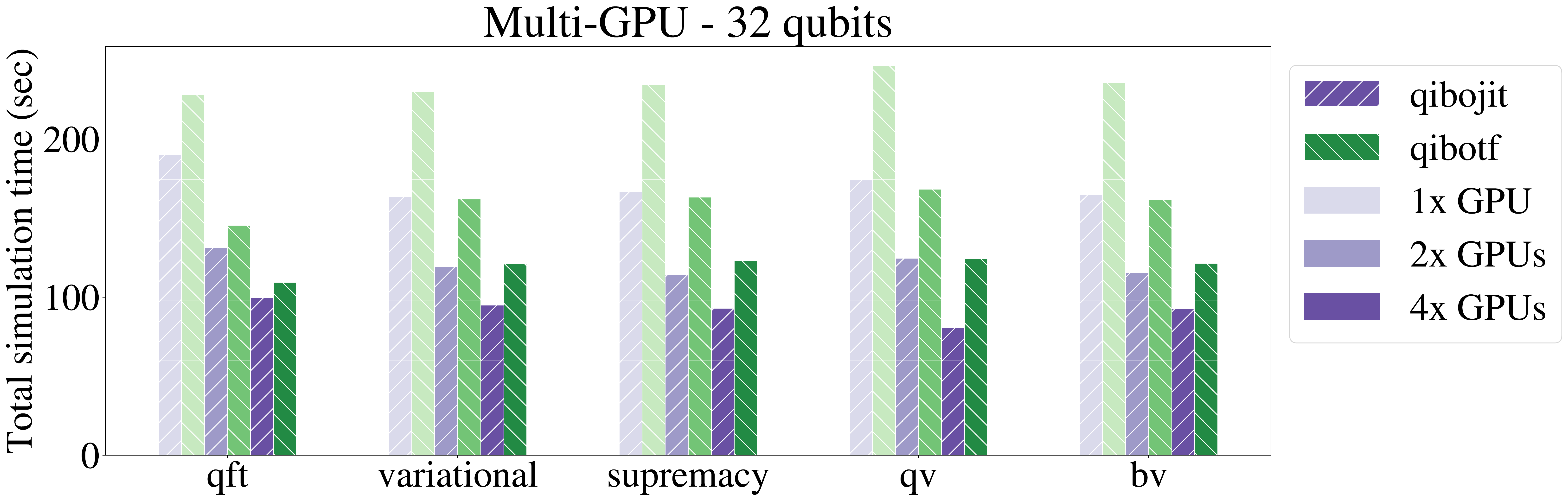}
    \caption{Total dry run (left) and simulation (right) time for simulating 32-qubit circuits using multiple GPUs.}
    \label{fig:multigpu}
\end{figure*}

To take full advantage of GPU acceleration for large circuits, Qibo provides the
possibility to simulate circuits on multiple GPU devices. This is useful because
the maximum number of qubits that can be simulated in a GPU is limited by its
internal memory. In Qibo's multi-gpu scheme, the full state vector is stored in
the host memory, which is typically larger than the GPU memory and only slices
of it are transferred to the GPUs for calculation.
For technical details of this implementation we refer to Sec.~2.5 Ref.~\cite{qibo2021}.
The multi-gpu scheme can be used with multiple
physical GPU devices, if available, but also with a single GPU that is re-used
for multiple state slices during the calculation. The multi-gpu feature is
supported by the {\tt qibojit} and {\tt qibotf} backends only.

In Fig.~\ref{fig:multigpu} we plot the times for simulating different circuits of 32 qubits using multiple
GPUs. The NVIDIA DGX workstation~\cite{dgx} with four Tesla V100 (32GB) GPUs, Intel Xeon E5-2698 CPU and 256GB of RAM
was used for this benchmark. For each circuit we distribute the execution to one, two or four physical GPUs.
The state slices are processed in parallel when different physical devices are used, while they are processed
sequentially if the same physical device is used. Therefore, increasing the number of physical devices leads
to better performance for both backends. We also observe that, even though for {\tt qibotf} there is no significant
variation between dry run and subsequent simulations, for {\tt qibojit} dry run appears slower, particularly when
multiple physical devices are used. This happens because parts of the calculation are not executed in parallel
in different devices during the just-in-time compilation step.

Finally, we performed comparisons with other open-source Python libraries for quantum simulation.
The libraries used are shown in Table~\ref{tab:library_versions} with their corresponding versions.
We focused on libraries that are compatible with multiple general-purpose, high-performance hardware
configurations, including multi-threading CPU and GPU. Some libraries allow switching between single
(complex64) and double (complex128) precision, while others support only a specific precision,
therefore we provide different comparisons for each case.
\begin{table}
\begin{tabular}{lccc}
    \toprule
    \textbf{Name} & \textbf{Version} & \textbf{Precision} & \textbf{Hardware} \\
    \midrule
    Qibo & 0.1.7 & single/double & CPU/GPU \\
    Qiskit & 0.34.2~\cite{qiskit_versions} & single/double & CPU/GPU \\
    Qulacs & 0.3.0 & double & CPU/GPU \\
    ProjectQ & 0.7.1 & double & CPU \\
    qsimcirq & 0.12.0 & single & CPU/GPU \\
    QCGPU & 0.1.1 & single & GPU \\
    HybridQ & 0.8.1 & single/double & CPU/GPU \\
    \bottomrule
\end{tabular}
\caption{Simulation libraries used in the benchmarks.}
\label{tab:library_versions}
\end{table}

\begin{figure*}
    \centering
    \includegraphics[width=0.41\textwidth]{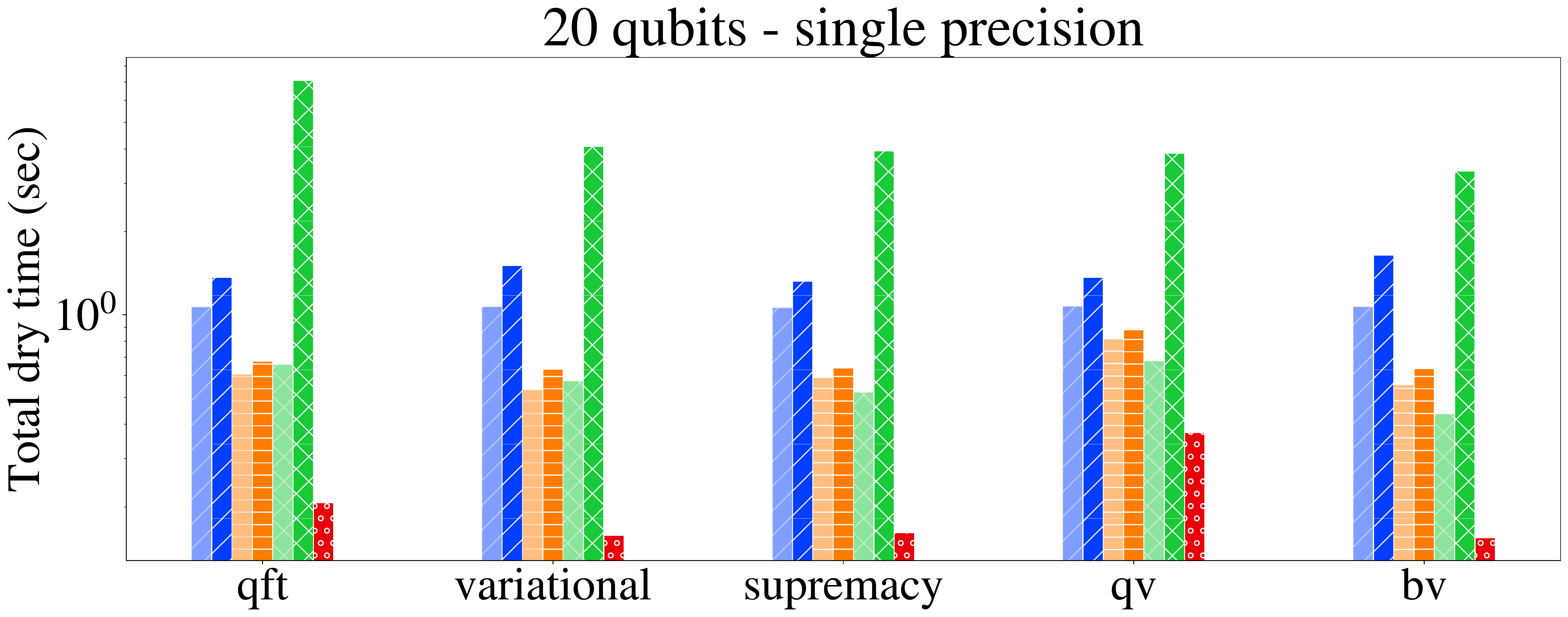}
    \includegraphics[width=0.5\textwidth]{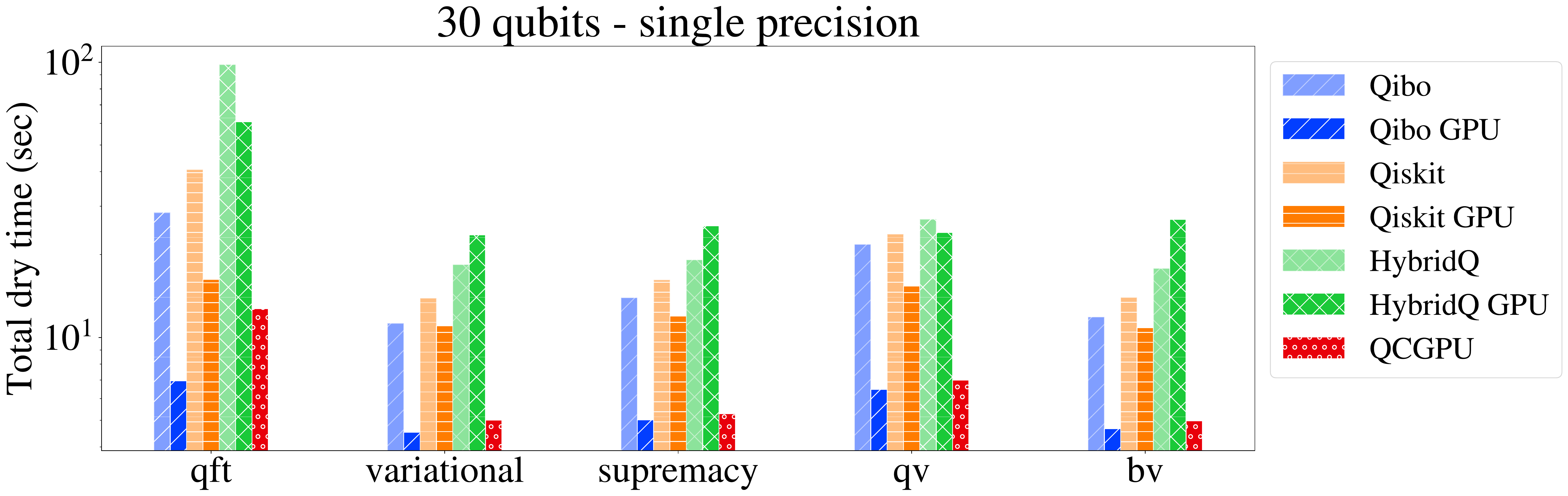}
    \includegraphics[width=0.41\textwidth]{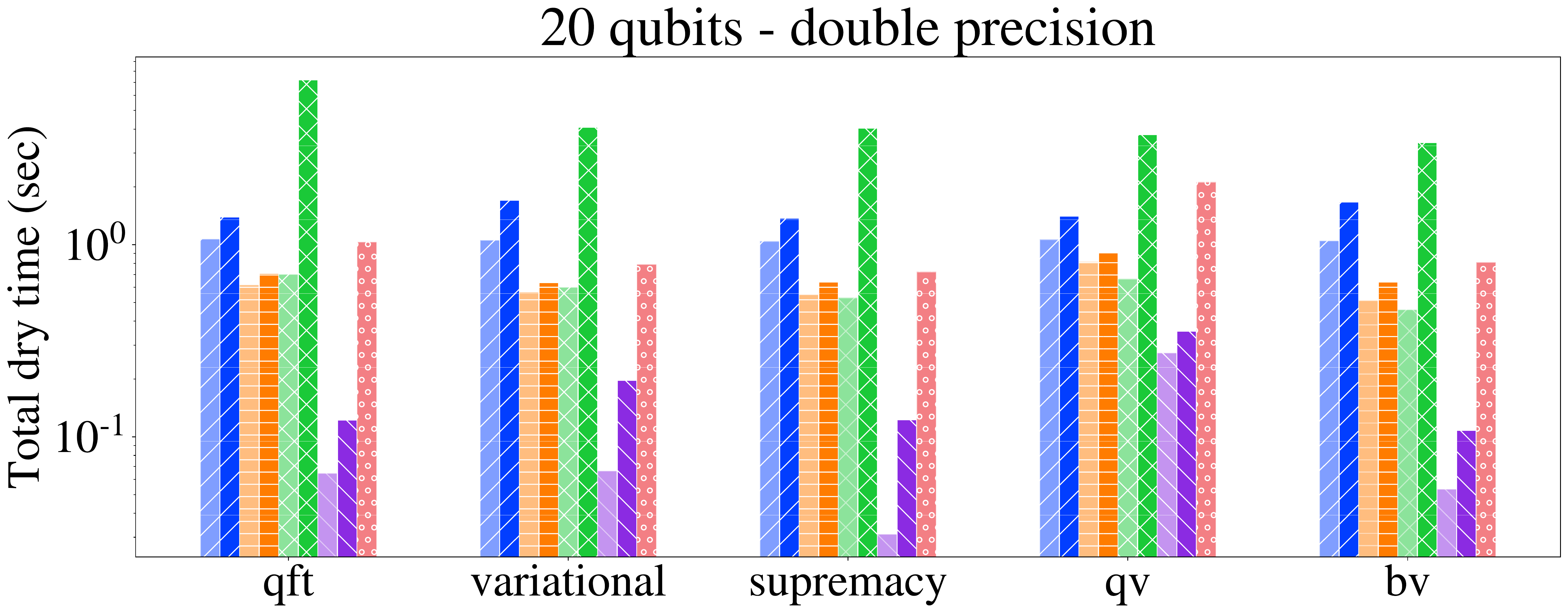}
    \includegraphics[width=0.5\textwidth]{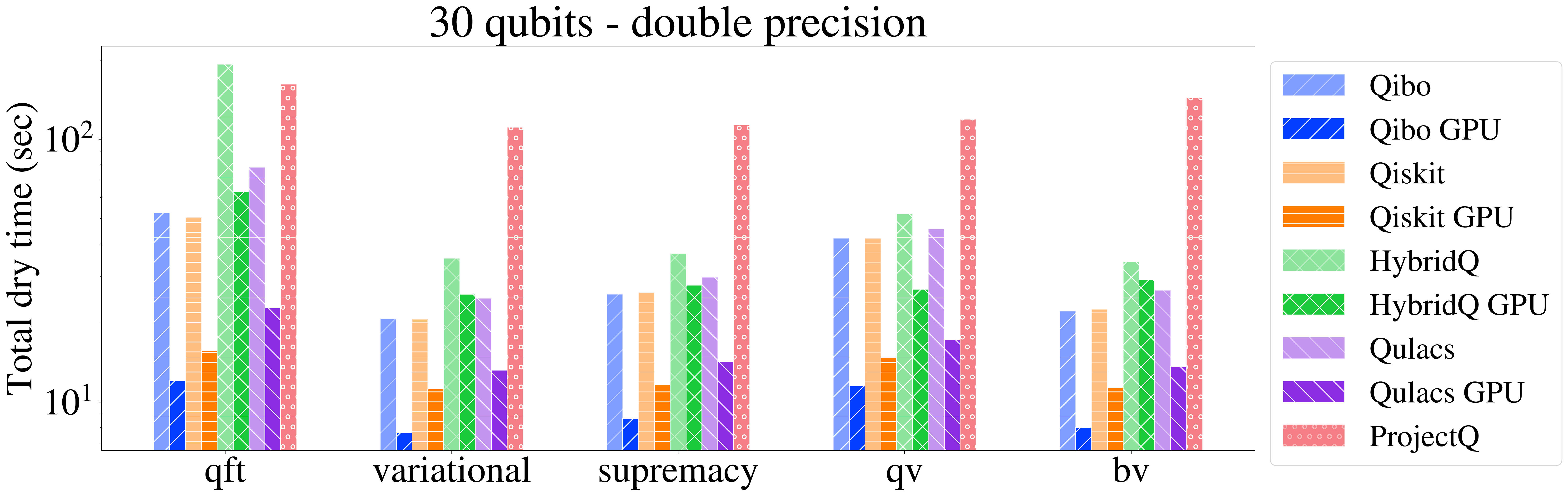}
    \caption{Total dry run time for simulating different circuits of 20 qubits (left) and 30 qubits (right), using libraries that support single (top) and double (bottom) precision.}
    \label{fig:libraries}
\end{figure*}

In Fig.~\ref{fig:libraries} we plot the comparison with different simulation libraries.
In each case we use all libraries that support that precision, and we benchmark each circuit from
Table~\ref{tab:benchmark_circuits} for 20 and 30 qubits on all supported hardware configurations
(multi-threading CPU and GPU). Optimizations such as gate fusion were disabled on all libraries
for this benchmark and the {\tt qibojit} backend (numba/cupy) was used for Qibo.
We do not include results for qsimcirq in this section, as it is not possible to disable gate
fusion for this library. We find that Qibo is slightly slower,
though still competitive when compared to other libraries, for circuits of 20 qubits.
This is primarily due to the import time and the time required to load the
kernels from disk, associated with just-in-time compilation.
These times are comparable to execution time for small circuits.
The situation is reversed for 30-qubit circuits where kernel loading times
are less relevant and Qibo is considerably fast, particularly on GPU.

\subsection{Gate Fusion}
\label{sec:gate_fusion_benchmarks}
Gate fusion~\cite{qhipster, tensorflow_quantum, qsim_arxiv} is a commonly used approach to speed up
simulation of quantum circuits.
Multiple gates are fused together by multiplying their underlying matrices and applying them
to the state vector as a single gate. This is preferable to naively applying the gates
one-by-one, particularly when simulating circuits with many qubits,
because multiplying small gate matrices is computationally cheaper than multiplying a gate matrix
with the exponentially large state vector.
Qibo provides a simple algorithm for fusing gates up to two target qubits.
This works by iterating over the circuit gates and greedily combining one-qubit and two-qubit gates
that act on the same target qubits.
\begin{figure*}
    \centering
    \includegraphics[width=0.58\textwidth]{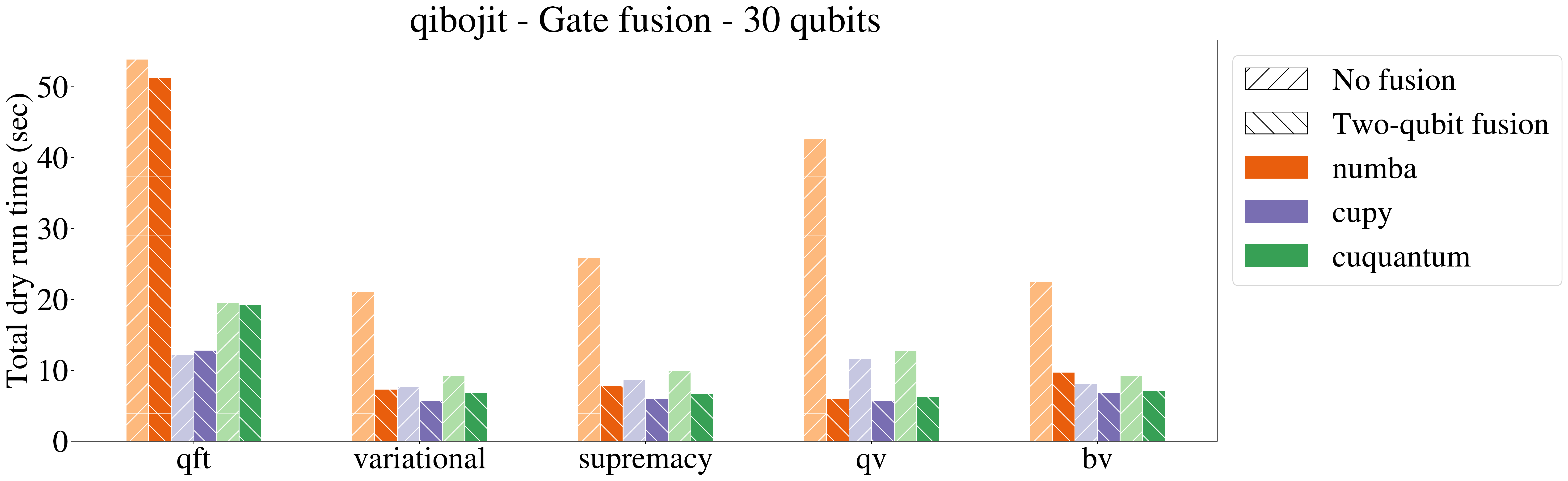}
    \caption{Dry run time for simulating different circuits of 30 qubits using qibojit with and without gate fusion.}
    \label{fig:qibojit_fusion}
\end{figure*}
\begin{figure*}
    \centering
    \includegraphics[width=0.6\textwidth]{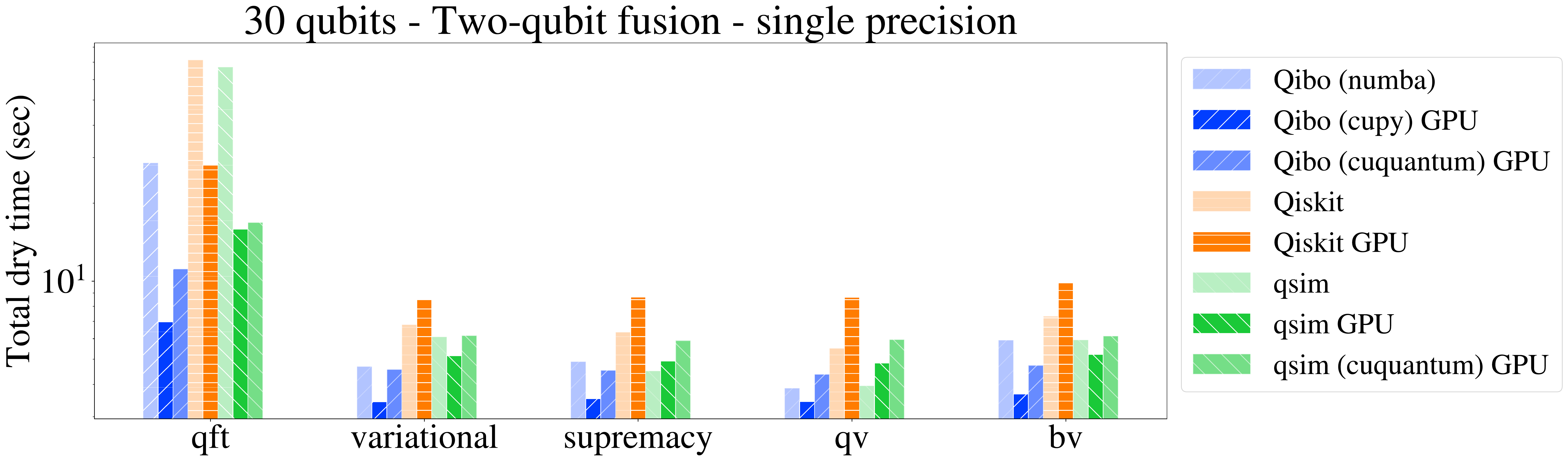}
    \caption{Dry run time for simulating different circuits of 30 qubits using different simulation libraries with gate fusion enabled.}
    \label{fig:libraries_fusion}
\end{figure*}

In Fig.~\ref{fig:qibojit_fusion} we compare all {\tt qibojit} platforms for simulating different
30-qubit circuits with and without fusion. The depth and the number of gates before and after the fusion
are shown in Table~\ref{tab:benchmark_circuits}. Fusion provides significant speed-up,
particularly when using CPU. However, it is important to note that this speed-up depends
on the circuit that is simulated. For example, gate fusion does not help much in the qft circuit.
In Fig.~\ref{fig:libraries_fusion} we compare different libraries on simulating different circuits
with gate fusion enabled. The maximum number of target qubits for a fused gate was set to two for
all libraries, in order to be consistent with the fusion algorithm used in Qibo.
Other libraries support fusion with higher maximum number of target qubits.
The most significant advantage appears when switching from no fusion to two-qubit fusion.
Further increasing the maximum number of qubits in fused gates up to about five,
may provide additional advantage for some circuits.

\subsection{Adiabatic Evolution}
\label{sec:evolution_benchmarks}
\begin{figure}
    \centering
    \includegraphics[width=0.43\textwidth]{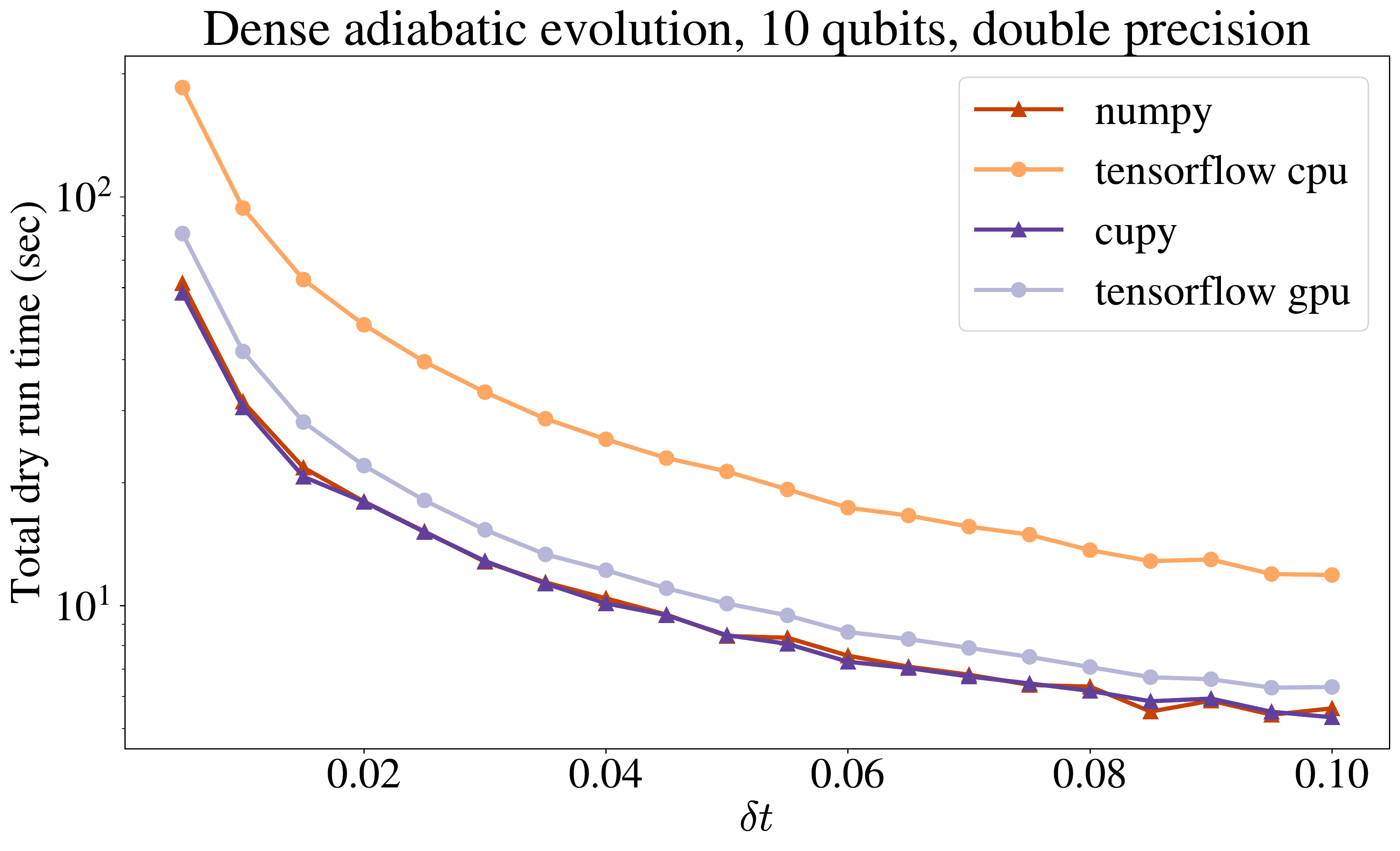}
    \caption{Total dry run time for simulating adiabatic evolution of 10 qubits using the full Hamiltonian matrix.}
    \label{fig:adiabatic_dense_dt}
\end{figure}
\begin{figure}
    \centering
    \includegraphics[width=0.44\textwidth]{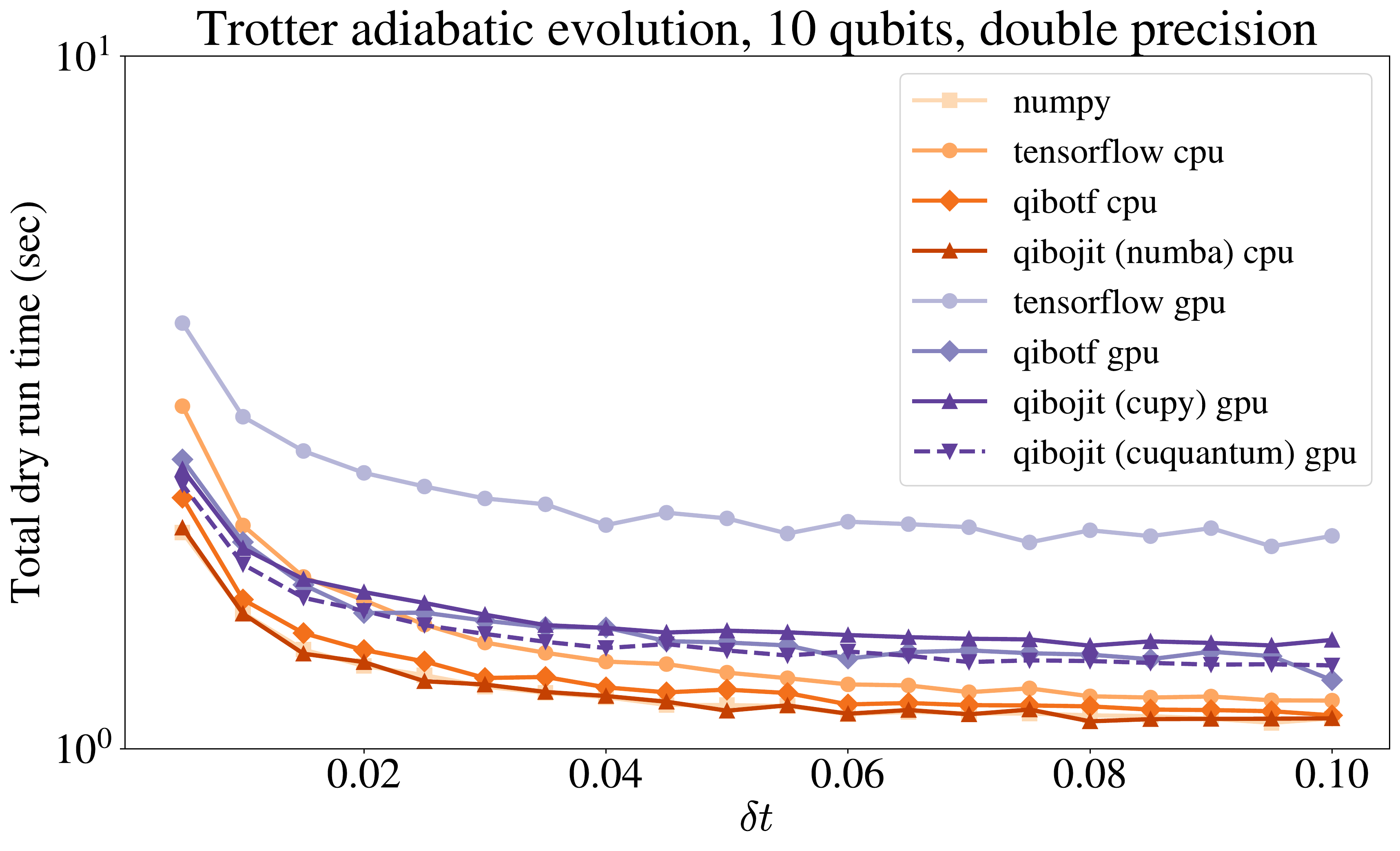}
    \includegraphics[width=0.44\textwidth]{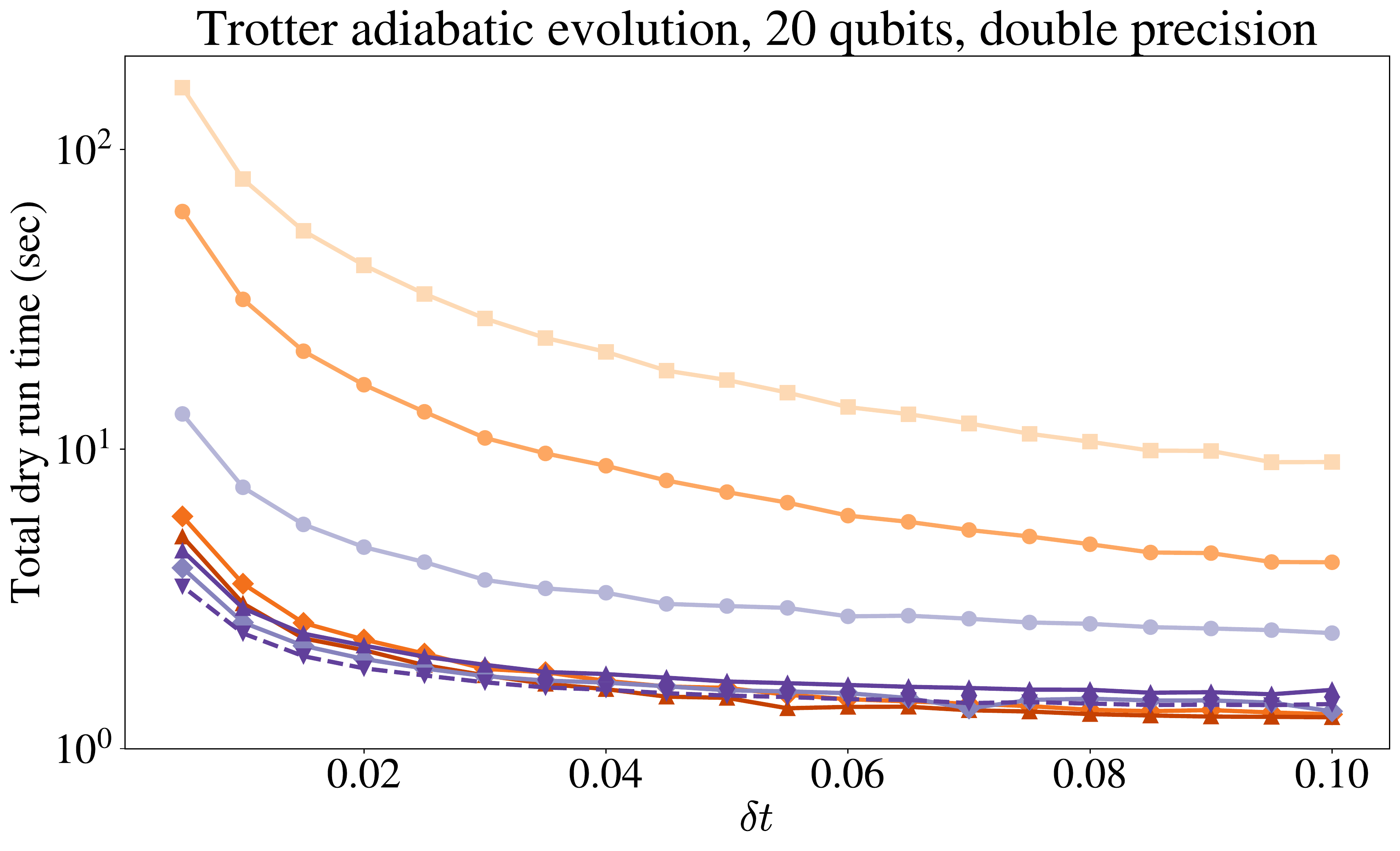}
    \caption{Total dry run time for simulating adiabatic evolution of 10 qubits (top) and 20 qubits (bottom) using the Trotter decomposition.}
    \label{fig:adiabatic_trotter_dt}
\end{figure}

Qibo provides functionality for simulation of unitary time evolution under
arbitrary Hamiltonians. A special case is adiabatic evolution, a typical method
for finding ground states of Hamiltonians~\cite{Kadowaki_1998,Crosson_2016},
which is provided as a special Qibo model. The trivial algorithm for unitary
time evolution calculates the exponential of the Hamiltonian matrix
$e^{-iH(t)\delta t}$ at each time $t$ where $\delta t$ is a pre-defined time
step. This approach is not feasible for large systems as the Hamiltonian matrix
for $n$ qubits has size $2^n\times 2^n$. For such cases, an alternative approach
based on the Trotter decomposition~\cite{Paeckel:2019yjf} is provided in Qibo
and can be used out-of-the-box, with the decomposition being handled
automatically by the library.

Here we benchmark the adiabatic evolution with the transverse-field Ising model (TFIM) as the target
Hamiltonian, starting from the easy to prepare Hamiltonian that is sum of Pauli X operators. We use
a system of 10 qubits and plot the scaling of execution time with the time step $\delta t$ used in
evolution, using both the matrix exponentiation (Fig.~\ref{fig:adiabatic_dense_dt}) and
Trotter decomposition (Fig.~\ref{fig:adiabatic_trotter_dt} top) methods.
With the Trotter decomposition we can also simulate a 20-qubit system
(Fig.~\ref{fig:adiabatic_trotter_dt} bottom) which is intractable when
using the full Hamiltonian matrix exponentiation.
As expected, the full exponentiation is computationally heavier and requires more time.
Moreover, this approach does not make use of custom operators but is based on numpy (CPU) and
cupy (GPU) primitives for the {\tt qibojit} backend and tensorflow (CPU and GPU) primitives
for the tensorflow and qibotf backends.
In contrast, when the Trotter decomposition is used, the time evolution is decomposed into
a circuit of unitary gates and all functionalities presented in Sec.~\ref{sec:circuit_benchmarks}
can be used. CPU is faster than GPU when simulating 10 qubits, however the
situation is reversed for 20 qubits, similar to what we observed in circuit simulation.

\section{Conclusion}
\label{sec:conclusion}

In this work we present the implementation of {\tt qibojit}, a just-in-time
compiled quantum simulator module for Qibo, with support on multi-threading CPU
and hardware accelerators (GPU and multi-GPU). We show that the modular
backend agnostic layout provided by Qibo simplifies the inclusion of new modules
with minor costs in terms of development time and maintainability.

Following the benchmark results presented in Sec.~\ref{sec:benchmarks} we can
confirm that {\tt qibojit} performance is acceptable and the impact of dry run
is negligible in most cases. The possibility to share the state vector
representation with external libraries such as cuQuantum, enhances furthermore
the capabilities of this module by allowing to obtain immediate performance
benefits from external specialized implementations.

In the short term, for quantum simulation we plan to explore the implementation
of alternative techniques to state vector simulation, while for QPU support we
are testing the framework on multiple real quantum hardware configurations.

\acknowledgments

We thank the NVIDIA Corporation team for supporting this project. The authors
would like to thank Christian Hundt for technical discussions about GPU
technology. We thank the Qibo team members for testing the code and providing
feedback concerning the results presented in this manuscript. S.C.~thanks Sofia
Vallecorsa and CERN's QTI for granting access to the ATOS QLM hardware.

\bibliographystyle{apsrev4-2}
\bibliography{references}

\end{document}